# An Overview of Privacy Dimensions on Industrial Internet of Things (IIoT)

Vasiliki Demertzi[1], Stavros Demertzis[2], Konstantinos Demertzis[3*]

[1]Computer Science Department, School of Science International Hellenic University Kavala Campus, 65404 Kavala, Greece; vademer@teiemt.gr
[2]School of Spatial Planning and Development Faculty of Engineering, Aristotle University of Thessaloniki 54124 Thessaloniki, Greece; demertzs@plandevel.auth.gr
[3*]School of Science & Technology, Informatics Studies, Hellenic Open University, Greece; demertzis.konstantinos@ac.eap.gr;

**Abstract:** Thanks to rapid technological developments, new innovative solutions and practical applications of the Industrial Internet of Things (IIoT) are being created, upgrading the structures of many industrial enterprises. IIoT brings the physical and digital environment together with minimal human intervention and profoundly transforms the economy and modern business. Data flowing through IIoT feed artificial intelligence tools, which perform intelligent functions such as performance tuning of interconnected machines, error correction, and preventive maintenance. However, IIoT deployments are vulnerable to sophisticated security threats at various levels of the connectivity and communications infrastructure they incorporate. The complex and often heterogeneous nature of chaotic IIoT infrastructures means that availability, confidentiality and integrity are difficult to guarantee. This can lead to potential mistrust of network operations, concerns about privacy breaches or loss of vital personal data and sensitive information of network end-users. This paper examines the privacy requirements of an IIoT ecosystem in industry standards. Specifically, it describes the industry privacy dimensions of the protection of natural persons through the processing of personal data by competent authorities for the prevention, investigation, detection or prosecution of criminal offences or the execution of criminal penalties. In addition, it presents an overview of the state-of-the-art methodologies and solutions for industrial privacy threats. Finally, it analyses the privacy requirements and suggestions for an ideal secure and private IIoT environment.

## Introduction

In a world where humans and machines are called upon to collaborate and coexist, the business organizational structures of the past create barriers and obstacles that not only waste energy unnecessarily but are likely to devalue information and hinder the diffusion of knowledge. At this point, the IIoT is transforming the way businesses operate and, by extension, those of the manufacturing industry [1]. IIoT is essentially about connectivity, in simple terms allowing objects, machines and devices to transfer data over a network without using human resources. In the future, it is possible that almost every manufactured item - whether infrastructure or consumable - will be embedded with sensors, allowing businesses to monitor performance and support failing operations, thereby adding value to the production process or journey of the buyer [2].

In this context, IIoT in industrial production is one of the most future-oriented technologies for industry because it combines two digitization strategies, Artificial Intelligence (AI) and Industry 4.0 [3, p. 4]. The product of these two technologies combines big data technologies with sensor connectivity and high-level automation technologies, which upgrade substantial the industrial environment [4].

On the other hand, this closer networking of the digital world of machines creates the potential for profound changes in the global industry and many areas of private and social life. Based on all this, it is necessary to present tomorrow's future trends in everything related to IIoT technology applications [5].

1. Growth of IIoT applications. Manufacturing automation continues to grow, with the number of companies choosing to automate and implement IIoT soaring to new levels due to the impact of the COVID-19 pandemic. Machine learning and robotics are two applications that increase automation. Machine learning increasingly automates manufacturing processes, so less human intervention is required, while the increasing number of human jobs being taken over by robotics results in fewer people in the workplace.
2. The wireless revolution. Not every IIoT application has access to local sockets. This is one of the reasons why more and more companies are using advanced IIoT wireless technologies, such as 5G, to connect to IIoT equipment in transit. The arrival of 5G is tailored to IIoT connectivity needs and makes a big change in connectivity for the industry in terms of Industry 4.0. An added advantage is that 5G ensures that the company's network is completely isolated, ensuring everything is secure and accessible.
3. Adoption of virtuaRealityty for remote operations. VirtuaRealityty becomes dominant for industrial applications regarding training and commissioning. Devices that combine a screen, camera and microphone become more sophisticated, and machine suppliers more often collaborate with their customers or service engineers through VR. The ability to commission machines remotely has made companies realize that being on-site is not always necessary. The machine supplier can work with the customer through an augmented reality headset such as a HoloLens. The customer sees virtual reality instructions and maintenance data to perform the necessary tasks, while the machine supplier receives a live feed of what the customer sees.
4. Use of machine data to improve customer relations. Connected machines have opened new ways to use machine data and improve customer relationships. It is not only interesting for large companies but also for smaller companies to make use of their data. Due to the increase in connected machines, the number of companies with access to critical machine data has also increased tremendously. It is a big challenge for many companies to discover new possibilities. The use of data is not only important to improve and optimize companies' machines, but also to create a better long-term relationship with customers. Machine data can, for example, be used to prevent equipment failures by predicting and performing machine maintenance before a fault occurs. In this way, machine downtime can ultimately be reduced [6].
5. Machine learning. Machine learning is a branch of AI where systems must be able to learn automatically and improve from experience without being programmed by humans. Applying machine learning can be quite difficult because preprocessing to label and normalize a lot of data takes time. Unsupervised learning or self-learning methodologies create higher-scale automation [7]. This means that human intervention is no longer needed since the data from the device is automatically sent to the algorithm. Thus, machine learning detects patterns of normal usage; therefore, after some time, it also tracks unusual patterns. For example, a machine creates several terns, but when a part of the machine fails, new patterns are created with donations from the normal pattern. When such a situation occurs, machinery suppliers receive a notification so they know that maintenance is required [8].
6. "Smart" packaging. Smart packaging using direct materials with built-in connectivity provides advanced benefits for industries. A primary aspect of smart packaging is that it enables consumers to engage with it and generate data to handle a product more effectively. Smart packaging can take the form of video recipes and other demonstrations that explain the use of the product. IIoT and packaging work together in different ways, including sensors, QR codes, and augmented reality/virtual reality/mixed reality options. The idea is to add consumer value and collect data through intelligent monitoring to optimize operations and enhance efficiency [9].

As can be easily seen, the development of IIoT is a big step in the realization of Industry 4.0 and the upcoming Industry 5.0, as it promotes the large-scale automation and optimization of processes related to intelligent sensors (e.g., configuration, high-volume handling data, decision-making, etc.). But this involves significant technical difficulties due to industrial wireless networks' large scale and complex structure. In addition, recording and transmitting large amounts of data create serious security and privacy concerns, as some may contain sensitive industry and personal information [10], [11].

## Dimensions of Privacy

Privacy in the industrial sector is a concept that is very difficult to define, especially in the digital age of IIoT, where the convergence of services creates unclear boundaries of definition. It can have many connotations depending on the contexts, relationships, and even products involved. To properly design privacy settings for IoT architectures, technologists must research and understand the dimensions of privacy that are important to the users of the services in question [12].

Privacy, as derived from EU directives and based on the way it was described by Martınez-Ballester et al. [13], can be categorized or take the following dimensions:

1. Identity Privacy. It concerns the identity details of an entity and is related to the concepts of authentication and authorization. Most of the data collected by the IIoT is intended for use by limited user groups [14]. Therefore, authentication (understanding the identity of the node or user) and authorization (by granting the necessary access permissions) are necessary, especially when it comes to issues of copyright, patents, etc., which are important issues for the existence and viability of an industry [15].
2. Location Privacy. It refers to an entity's location identification information. Said determination violates personal or industrial privacy issues concerning the detection, identification, storage, processing and sharing of information in a technical or legal context [16].
3. Footprint Privacy. It refers to an entity's unique traceable communications actions. A feature of this function can be found in Smart Energy Grids, characterized by real-time two-way communications [17]. How to control and safely retrieve energy data shared with third parties poses a challenge to the privacy of network users. A robust solution towards solving problems related to Footprint Privacy is the PaRQ [18] standard proposed by Wen et al., which allows a home user to store measurement data on a cloud server in encrypted form. When financial audits are required, an authorized requester can send a pair of queries to the cloud server to retrieve the measurement data.
4. Multidimensional Privacy. It refers to an entity's multidimensional or complex identification elements, which may combine some of the above dimensions [13]. Solving such problems requires complex combined processes or solving and parameterizing the individual issues in a custom schema [19].

It should be noted that no relevant regulation exists for protecting natural persons' privacy, exclusively for the industrial domain. However, almost every industry is involved in processing personal data in one or more processes. For the processing of personal data in the industrial domain and the free movement of such data by competent industrial authorities for the prevention, investigation, detection, or prosecution of criminal offences or the execution of sanctions, industries follow the existing regulators such as the GDPR. In this regard, the following industries were impacted [20]:

1. Employee personal data processing: It isn't easy to imagine a business without employees. That is, all organisations employ people. Employees are also data subjects who are subject to regulatory privacy policies. As a result, organisations must be more transparent and accountable when processing their employees' data.
2. Processing sales contacts' data: It isn't easy to imagine a business without customers. The very existence of an organisation is to serve customers. The contacts are real people, even if a company's customer is another company. Making sales contacts, maintaining their data, and so on are all activities that would be classified as processing personal data. As a result, organisations across all industries must ensure that this processing adheres to privacy rules and regulations.
3. A Data Protection Officer (DPO) must be appointed. A DPO is usually appointed by organisations that process large amounts of personal data. This will occur in all industries to comply with privacy regulations.

**Industrial Privacy**

The concept of privacy, as attributed by Boussada et al. [21], is the right of individuals to control or influence what information related to them may be collected and stored and by whom and to whom that information may be disclosed. The need to protect privacy is highlighted more strongly when the quantitative and qualitative difference in the possibility of collecting and processing information is perceived, as in the IIot environment. These possibilities, as highlighted by the Industrial Internet Consortium in Vol 4 [22], make the issue of privacy a very important factor, as the multi-functional use and the decentralization of information from its original carrier, as it is applied in IIoT environments, creates serious issues privacy, related to its collection, processing and final disposal.

Privacy and data protection are two of the most pressing issues facing businesses in the industry. Most consumers regard personal information as particularly sensitive, especially personal financial data. The respective regulatory authorities promote good data management practices to increase customer profiling to identify potential opportunities and conduct risk management analysis. To that end, managing privacy and data protection are critical throughout the customer lifecycle. Several use cases in the industrial sector, for example, involve data sharing between different organizations (e.g. data for customer protection or faster transactions, business data sharing for improved credit risk assessment, customer or worker insurance data sharing for faster claims management and others) [23], [24].

Leaders in industries such as manufacturing or heavy industry do not rank data privacy at the top of their organization's list of concerns. The reason is that they consider that the privacy of the data they manage is not as important as other companies, such as retail companies, financial institutions and industrial healthcare systems, where personal data has a different value. However, the space is changing in the digital age, and data privacy is a key risk for any industry that handles potentially sensitive data related to its customers, employees and business partners [25].

Considering the heterogeneity of the systems included in an IIoT ecosystem and the non-statutory interoperability at the hardware and software level, serious objections arise to securing these systems. In the IoT ecosystem, where various systems interact with the physical world, the uncontrolled arrangement of states can lead to dangerous conditions, especially where these systems have data flow from multiple intermediaries, requiring multi-layered security approaches beyond link encryption. More generally, the convergence of computation and communication technologies, decentralization of processing, distributed analysis, interconnection, and data sharing in every industrial activity radically change the concept of privacy and generate serious new challenges [26], [27].

In this context and in contrast to privacy in its narrow sense, the protection of industrial privacy is raised as a primary demand linked to technological processes and the use of IIoT, as it is assessed that the existing regulations - standardization, do not offer a robust shield against the looming dangers. For example, traditional Omnidirectional Antennas and MAC-Access Control (MAC) protocols cannot be used in 5G systems, so more sophisticated and perhaps more complex privacy solutions should be offered, such as the one proposed by Szymanski [28]. This proposal demonstrates that combining a major Software-Defined Network (SDN) control layer, low-spin deterministic scheduling, and lightweight encryption at Layer 2 can provide a new approach to wireless security and performance with multiple capabilities for privacy-preserving IIoT standardization systems [11].

The conclusion of insufficient privacy protection in IIoT systems results in both from the heterogeneity of interconnected industrial infrastructures and the exponential increase in the level of sophistication of cyber-attacks such as Targeted Ransomware and Hijacked Two-Factor, which redefine the need to review the security of IIoT subsystems. This need, combined with the redefinition and evaluation of security controls for protecting privacy in IIoT architectures and the determination of more strict policies for analyzing the effectiveness of specific security and privacy controls [29] in said environments and how they are applied in the design and development of new IIoT systems, suggest Hassanzadeh et al [30].

Correspondingly, it is also vital to upgrade or adapt the already existing standards – systems promoted in industrial technologies, as in the case of Occhiuzzi et al. [31], who applied the Radiofrequency Identification of the emerging applications to the low-level monitoring of critical infrastructures to detect early attempts at physical and cyber-attacks. At the same time, there should be continual updating and redefinition of the security requirements of IIoT systems with the control of certification and identity, such as RFID systems [32]. Also, a very important role in ensuring privacy in the IIoT environment is the knowledge of existing best practices and recommendations for maintaining security and future directions for continuous improvement and adaptation. Considering the particular capabilities and processing of information related to the ICT ecosystem, even seemingly "harmless" information has its informational utility. Their final value is determined by their processing, their combination and the environment in which they are reported and evaluated [33].

# Privacy Threats in the IIoT

Adopting increasingly powerful and complex IP-based devices, such as sophisticated microprocessors, is challenging to face cyber security and privacy issues [34]. From manufacturing to health care, IIoT systems improve service delivery and increase productivity. However, IIoT devices are vulnerable like anything else connected to the internet [35]. Attacks on industrial control systems (ICS) such as distributed control systems (DCS), programmable logic controllers (PLC), supervisory control and data acquisition (SCADA), and human-machine interfaces are examples (HMI) [36].

A comprehensive set of security and privacy solutions that do not disrupt operations, service reliability, or profitability should be used to protect IIoT infrastructure [37]. A practical, simple, yet secure solution that IIoT device manufacturers and their customers can easily and widely adopt is more effective than a super solution' that fails to gain significant traction [38]. The following capabilities should be included by design to reduce significant privacy threats in the IIoT [37], [39]:

1. Firmware integrity and boot security. Secure boot employs cryptographic code signing techniques to ensure that a device executes only code generated by the device's OEM or another trusted party. By utilizing secure boot technology, hackers are prevented from replacing the firmware with malicious instruction sets, thereby preventing attacks. Unfortunately, not all IIoT chipsets include secure boot support. In such a case, it is critical to ensure that IIoT devices can only communicate with authorized services to avoid the risk of malicious instruction sets replacing the firmware [40].
2. Mutual identification. Before receiving or transmitting data, a smart actuator on the manufacturing floor should be authenticated every time it connects to the network. This ensures that the data comes from a legitimate device rather than a fraudulent source. Secure mutual authentication, in which two entities (device and service) must prove their identities to each other, aids in the prevention of malicious attacks. Cryptographic algorithms with symmetric or asymmetric keys can be used for two-way authentication. The Secure Hash Algorithm (SHA-x) and hash-based message authenticated code (HMAC) can be used for symmetric keys. In contrast, Elliptic Curve Digital Signature Algorithm (ECDSA) can be used for asymmetric keys [41].
3. Secure communication (end-to-end encryption). Data in transit between a device and its service infrastructure is protected by secure communication capabilities (the cloud). A smart actuator, for example, that sends usage data to the SCADA must be able to protect information from digital eavesdropping. Encryption ensures that only those witnesses to a secret decryption key can access transmitted data. Encryption ensures that only those witnesses to a secret decryption key can access transmitted data [3].
4. Monitoring and analysis of security. Data on the overall state of an industrial system, including endpoint devices and connectivity traffic, is captured by security monitoring. The data is then analyzed to detect potential security violations or system threats. It is critical to protect endpoint devices from tampering and data manipulation, which could result in inaccurate event reporting. When abnormal behaviour is detected [42], [43], a wide range of actions should be taken as part of an overall system security policy, such as revoking device credentials or quarantining an IoT device. This automatic monitor-analyze-act cycle can be run in real-time or later to identify usage patterns and potential attack scenarios [44].
5. Management of the security lifecycle. The lifecycle management feature enables service providers and OEMs to control IoT devices' security while in use. Rapid over-the-air (OTA) device key(s) replacement during cyber disaster recovery ensures minimal service disruption. Secure device decommissioning also ensures that scrapped devices are not repurposed and used to connect to a service without authorization [45].

The main concerns regarding protecting data privacy in industrial domains are intertwined with the more general concerns about the risks associated with each modern network device [46]. In general, the most basic and common privacy threats related to IIoT are:

1. Identification and Authorization. It is directly related to the concept of Identity Privacy. It refers to the effort to find correlations between data that can be used to detect, identify and maliciously replicate the application of profiles (sets of associated data) to personalize and identify secret, industrial information. Techniques such as Subscriber Identity Module (SIM) and Machine Identification Module (MIM), proposed by Borgia [14], are important solutions worthy of attention. However, these approaches work in centralized single-management networks. At the same time, in distributed topologies, it isn't easy to manage identification services and standardizations such as the one proposed by Moosavi et al [15]. and concerns an architecture of authorization of remote end users by distributed smart gateways, which are based on the Datagram Transport Layer Security (DTLS) handshake protocol. In addition, IIoT attacks compromise authorized industrial systems access, and as a result, one such security issue can degrade the related services. Ransomware also causes IIoT devices to malfunction and steals users' sensitive information and data. In addition, if a large number of smart IIoT devices are unable to encrypt user data, malware will emerge [47]. To prevent unauthorized device access, IIoT devices use a network that does not convert data into code.
2. Localization and tracking. It is directly related to the concept of Location Privacy. An industry can choose the locations it chooses to perform its economic functions. Several issues influence the choice of a suitable location, most importantly the nature and characteristics of the industrial activity carried out by the enterprise (e.g. extraction of raw materials or cultivation, production of intermediate or final products, provision of a service) and the associated costs of production, balanced with the cost of physical distribution to target markets and the importance of proximity to customers as a basis for establishing competitive advantages over rival suppliers. Similarly, many service activities must be located in and around the customer's catchment areas. At the same time, some suppliers may be interested in operating alongside their core customers to synchronize production input requirements better. Some locations may be preferred for their production advantages, for example, due to lower labour costs or the availability of investment subsidies or the supply of skilled workers and parallel access to relevant facilities. On the other hand, the high cost of distribution, especially in the case of bulky products with low added value or the international context, the imposition of tariffs and quotas on imports, creates important requirements for an appropriate position oriented to the market but protected from the prying eyes of the competition and espionage. A low-cost technical solution that adds protection to the IoT environment was proposed by Joy et al [16]. by embedding in GPS devices privacy software that ensures that IoT devices and their administrators have fine-grained control over releasing their position. In addition, the safety of data ingested from numerous IIoT devices is related directly to other data security and privacy concerns from insecure cloud infrastructures, web applications, and mobile environments. As a result, it is necessary to follow data transmission security rules in each environment so that there are measures in place to identify the path from whose device the data is transmitted. It is also critical to eliminate irrelevant data and data without relation to the actual operation. Although compliance with numerous regulatory structures becomes difficult when multiple data is stockpiled, the infrastructure must be carried out with separate services for controlling data linked to interconnected devices and environments [48].

3. Profiling. The specific threat lies in violating privacy and monitoring persons or individuals in their association with specific industrial processes. Accordingly, it may refer to the identification, collection and processing of information derived from services or reference models, which may constitute an industrial secret. Characteristics of the ongoing concern for protecting IoT devices from profiling threats are efforts to enhance privacy in RFID devices [49], [50], sensor systems [51] and wireless networking [52], [53], and identity management [50], [54] technologies to enhance privacy or encryption technology [55], [56].

4. Hardware Lifecycle. Industrial devices are, in most cases, remanufactured and reused. Therefore, sensitive information, device logs, and data stored in memories or storage media will likely fall into the wrong hands with unpredictable consequences [57]. For the specific threats, the industry should draw up and implement a uniform policy for the management of industrial equipment, as well as apply techniques of total deletion [58] of the data locally or distributed information processing systems which include first and second sites, which may include corresponding information production and copying sites [59]. Also, IIoT hardware addresses security and privacy threats from inadequate testing and a lack of upgrading processes [60]. IIoT device manufacturers, while willing to produce various devices, do not consider the security and upgrading concerns of said devices because they require extensive testing and, therefore, additional costs. These malfunctions increase the possibility of security and privacy attacks when released into a real-world industrial infrastructure [61].

5. Inventory attack. Inventory attacks refer to the unauthorized collection of information about the existence and characteristics of the equipment. Also, with the implementation of the M2M vision [62], [63], smart devices can, subject to conditions from any legal or hypothetically legal entities, be asked for information related to, for example, their energy footprint, communication speeds, reaction times, as well as other unique characteristics, which could potentially be used to identify their type and model. Thus, malicious users violating the privacy of an industry can compile an inventory list of the devices in a specific building or factory, along with information on how it works [64]. Here too, cryptography solutions have been proposed for aggregation mechanisms. This secure aggregation protocol meets the IoT requirements [65]. It analyses its efficiency considering various system configurations and the impact of the wireless channel through packet error rates and private communication mechanisms [66].

6. Linkage. This threat consists of connecting different previously separated systems so that the combination of the data and the sources reveals critical information that would be impossible to reveal by individual systems [37]. Moreover, to ensure the smooth operation of IIoT devices, it is critical to have flat networking that will allow them to function effectively. It is critical to have a high-quality open networking system for this purpose [67]. This particular factor in IIoT networks creates a security barrier. In this regard, industrial enterprises must thoroughly assess their security policies to ensure that IIoT devices are not vulnerable to threats [68]. Also, providers must understand the significance of properly configuring the networking device and services and that data privacy entails various processes, such as efficiently removing sensitive information through data segregation [48].

However, a significant part of the responsibility lies with the hardware manufacturers, as often, the mechanisms and interoperability standards concerning the security of IIoT devices are either neglected or treated as a secondary consideration [69]. Usually, this is due to the requirement for a short period to implement an IoT device, simplify the design of its operating mechanisms, and reduce its overall cost. It is therefore considered important that those involved in IIoT delivery processes consider privacy and develop privacy management interfaces built into the endpoint and web interface of the product or service [70]. This technology should allow the user-industry to determine which privacy features are used by the ecosystem, what the Terms of Service are, and that it is possible to disable the exposure of this information to the business, its partners or competitors. This details management system will help ensure that users have the right and ability to control the information they share about themselves and their physical world [71].

**Privacy Requirements and Suggestions**

Privacy requirements entail the application of all applicable privacy laws and all applicable industry policies, notices and contractual obligations regarding the collection, recording, use, storage, processing, sharing or sharing, protection, security (technical, physical and administrative), disposal, destruction, disclosure or transmission (including cross-border) of sensitive personal data [72]. But because something so strict in an industrial environment is practically impossible to implement, below are summarized the most basic requirements - recommendations that should be followed in an industrial environment that respects privacy:

1. Mitigation by Design. Privacy protection solutions should ideally be anticipated and incorporated during the design phase of IoT products, services or systems [73], [74].
2. Assessment. Privacy impact assessments should help provide a secure method of analyzing how and when personally identifiable information is collected, stored, protected, shared and managed, and how it is disposed of [75]–[77].
3. Legal Compliance. An assessment of applicable legal or regulatory requirements should be performed to monitor compliance [78]–[80].
4. Use Limitation. Provision should be made for the necessary work to ensure that access to any physical or electronic security system is restricted to fully authorized persons and for fully authorized purposes [81]–[84].
5. Storage Safeguards. Warehouses, data lakes, databases, where personal information is collected and stored, should be protected in terms of physical and logical security [85]–[88].
6. Secure Communications. Data transmitted between systems or components, and more generally communications in an IoT environment, should be protected from unauthorized disclosure or access [83], [89]–[92].
7. Transparency. Individuals whose personal information may be collected should be notified of the reason for collection and how that information may be used. There should also be mechanisms that can reveal possible leaks of personal data [93]–[95].
8. Data Retention Policy. There should be a policy that defines the retention period of personal data, the methods of destruction of such data and a procedure that ensures that deleted information is not recoverable [1], [67], [96]–[98].

To support the requirements above and recommendations, industries must reconsider data use and regulations to unlock the value of data while strengthening consumer trust and protecting their fundamental rights. A permissioned industrial infrastructure that offers privacy control, auditability, secure data sharing, and faster operations must be strengthened in two ways:

1) Including features and associated cryptography to facilitate personal asset improvements (e.g., sharing personal data) via a secure platform based on emerging technologies such as blockchain.
2) Employing techniques like Multi-Party Computation (MPC) [99], [100] and Linear Secret Sharing (LSS) [101] to allow searching of encrypted data as a means of providing higher data privacy guarantees.

Based on these suggestions, the industrial partners will enable disruptive business models for personalization and full automation of secure and private processes.

## Conclusion

The adoption of secure and privacy IIoT in Industry 4.0 solutions by industries has created several challenges in this sector. Integrating emerging secure technologies by businesses contributes to their further digital transformation, which is also a key challenge for the modern industry. The magnitude of this challenge is related to the degree of readiness and maturity of companies to integrate secure Industry technologies into their production process, both at the level of secure infrastructure and the level of privacy-preserving services. In addition to the digital transformation, the adoption of privacy-preserving IIoT technologies creates the foundations for Flexible Manufacturing (Agile Manufacturing), where companies will be able to react immediately to market changes since the production process with the use of new technologies will be connected to the supply chain as well as with end users. This interconnected production creates the need for new kinds of business models as well as the protection, control and management of data.

## Conflicts of Interest

The authors declare no conflict of interest.

## Data Availability Statement

The data used in this study are available from the author upon request.

## Funding Statement

This research received no external funding.

## References

[1]  I. Alqassem and D. Svetinovic, "A taxonomy of security and privacy requirements for the Internet of Things (IoT," in *2014 IEEE International Conference on Industrial Engineering and Engineering Management*, Bandar Sunway, 2014, pp. 1244–1248. doi: 10.1109/IEEM.2014.7058837.
[2]  R. Bogue, "Cloud robotics: a review of technologies, developments and applications," *Ind. Robot Int. J.*, vol. 44, no. 1, pp. 1–5, Jan. 2017, doi: 10.1108/IR-10-2016-0265.
[3]  S. Sulaiman, A. Aldeehani, M. Alhajji, and F. A. Aziz, "Development of integrated supply chain system in manufacturing industry," *J. Comput. Methods Sci. Eng.*, vol. 21, no. 3, pp. 599–611, Jan. 2021, doi: 10.3233/JCM-200045.

[4] Y. Hui and L. Zesong, "Research on Real-time Analysis and Hybrid Encryption of Big Data," in *2019 2nd International Conference on Artificial Intelligence and Big Data (ICAIBD)*, Feb. 2019, pp. 52–55. doi: 10.1109/ICAIBD.2019.8836992.

[5] B. Hamid, N. Jhanjhi, M. Humayun, A. Khan, and A. Alsayat, "Cyber Security Issues and Challenges for Smart Cities: A survey," in *2019 13th International Conference on Mathematics, Actuarial Science, Computer Science and Statistics (MACS)*, Sep. 2019, pp. 1–7. doi: 10.1109/MACS48846.2019.9024768.

[6] P. Akubathini, S. Chouksey, and H. S. Satheesh, "Evaluation of Machine Learning approaches for resource constrained IIoT devices," in *2021 13th International Conference on Information Technology and Electrical Engineering (ICITEE)*, Jul. 2021, pp. 74–79. doi: 10.1109/ICITEE53064.2021.9611880.

[7] I. H. Sarker, "Deep Learning: A Comprehensive Overview on Techniques, Taxonomy, Applications and Research Directions," *SN Comput. Sci.*, vol. 2, no. 6, p. 420, Aug. 2021, doi: 10.1007/s42979-021-00815-1.

[8] M. Djibo, M. Y. El-Sharkh, and N. Sisworahardjo, "Fuzzy Artificial Immune System based Generators Preventive Maintenance Scheduling," in *SoutheastCon 2022*, Mar. 2022, pp. 649–654. doi: 10.1109/SoutheastCon48659.2022.9764085.

[9] N. Akhtar, A. Mian, N. Kardan, and M. Shah, "Advances in Adversarial Attacks and Defenses in Computer Vision: A Survey," *IEEE Access*, vol. 9, pp. 155161–155196, 2021, doi: 10.1109/ACCESS.2021.3127960.

[10] L. Dias and T. A. Rizzetti, "A Review of Privacy-Preserving Aggregation Schemes for Smart Grid," *IEEE Lat. Am. Trans.*, vol. 19, no. 7, pp. 1109–1120, Jul. 2021, doi: 10.1109/TLA.2021.9461839.

[11] G. Drosatos, K. Rantos, D. Karampatzakis, T. Lagkas, and P. Sarigiannidis, "Privacy-preserving solutions in the Industrial Internet of Things," in *2020 16th International Conference on Distributed Computing in Sensor Systems (DCOSS)*, Feb. 2020, pp. 219–226. doi: 10.1109/DCOSS49796.2020.00044.

[12] G. Guo, Y. Zhu, R. Yu, W. C.-C. Chu, and D. Ma, "A Privacy-Preserving Framework With Self-Governance and Permission Delegation in Online Social Networks," *IEEE Access*, vol. 8, pp. 157116–157129, 2020, doi: 10.1109/ACCESS.2020.3016041.

[13] A. Martınez-Balleste, P. A. Perez-Martınez, and A. Solanas, "The pursuit of citizens' privacy: a privacy-aware smart city is possible," *IEEE Commun. Mag.*, vol. 51, no. 6, 2013, doi: 10.1109/MCOM.2013.6525606.

[14] E. Borgia, "The Internet of Things vision : Key features , applications and open issues," *Comput. Commun.*, vol. 54, pp. 1-31, 2014.

[15] S. R. Moosavi *et al.*, "SEA: A Secure and Efficient Authentication and Authorization Architecture for IoT-Based Healthcare Using Smart Gateways," *Procedia Computer Science*, vol. 52. 2015. doi: 10.1016/j.procs.2015.05.013.

[16] J. Joy, M. Le, and M. Gerla, "LocationSafe: granular location privacy for IoT devices," in *Proceedings of the Eighth Wireless of the Students, by the Students, and for the Students Workshop*, New York, NY, USA, Jul. 2016, pp. 39–41. doi: 10.1145/2987354.2987365.

[17] K. Demertzis, L. S. Iliadis, and V.-D. Anezakis, "An innovative soft computing system for smart energy grids cybersecurity," *Adv. Build. Energy Res.*, vol. 12, no. 1, pp. 3–24, Jan. 2018, doi: 10.1080/17512549.2017.1325401.

[18] M. Wen, R. Lu, K. Zhang, J. Lei, X. Liang, and X. Shen, "PaRQ: A Privacy-Preserving Range Query Scheme Over Encrypted Metering Data for Smart Grid," *IEEE Trans. Emerg. Top. Comput.*, vol. 1, no. 1, pp. 178-191, Jun. 2013, doi: 10.1109/TETC.2013.2273889.

[19] R. D. Garcia, G. Sankar Ramachandran, R. Jurdak, and J. Ueyama, "A Blockchain-based Data Governance with Privacy and Provenance: a case study for e-Prescription," in *2022

*IEEE International Conference on Blockchain and Cryptocurrency (ICBC)*, Feb. 2022, pp. 1–5. doi: 10.1109/ICBC54727.2022.9805545.

[20]    C. Sullivan, "EU GDPR or APEC CBPR? A comparative analysis of the approach of the EU and APEC to cross border data transfers and protection of personal data in the IoT era," *Comput. Law Secur. Rev.*, vol. 35, no. 4, pp. 380–397, Aug. 2019, doi: 10.1016/j.clsr.2019.05.004.

[21]    R. Boussada, M. E. Elhdhili, and L. A. Saidane, "A survey on privacy: Terminology, mechanisms and attacks," in *2016 IEEE/ACS 13th International Conference of Computer Systems and Applications (AICCSA*, Agadir, 2016, pp. 1–7. doi: 10.1109/AICCSA.2016.7945804.

[22]    "Industrial Internet Security Framework," *Industry IoT Consortium*. https://www.iiconsortium.org/iisf/ (accessed Jan. 15, 2023).

[23]    R. Bonacin, M. Fugini, R. Martoglia, O. Nabuco, and F. Saïs, "Web2Touch 2020–21 : Semantic Technologies for Smart Information Sharing and Web Collaboration," in *2020 IEEE 29th International Conference on Enabling Technologies: Infrastructure for Collaborative Enterprises (WETICE)*, Sep. 2020, pp. 235–238. doi: 10.1109/WETICE49692.2020.00053.

[24]    S. A. Butt, J. L. Diaz-Martinez, T. Jamal, A. Ali, E. De-La-Hoz-Franco, and M. Shoaib, "IoT Smart Health Security Threats," in *2019 19th International Conference on Computational Science and Its Applications (ICCSA)*, Jul. 2019, pp. 26–31. doi: 10.1109/ICCSA.2019.000-8.

[25]    R. Duezguen *et al.*, "How to Increase Smart Home Security and Privacy Risk Perception," in *2021 IEEE 20th International Conference on Trust, Security and Privacy in Computing and Communications (TrustCom)*, Jul. 2021, pp. 997–1004. doi: 10.1109/TrustCom53373.2021.00138.

[26]    S. Aich, S. Chakraborty, M. Sain, H. Lee, and H.-C. Kim, "A Review on Benefits of IoT Integrated Blockchain based Supply Chain Management Implementations across Different Sectors with Case Study," in *2019 21st International Conference on Advanced Communication Technology (ICACT)*, Oct. 2019, pp. 138–141. doi: 10.23919/ICACT.2019.8701910.

[27]    H. K. Sharma, A. Kumar, S. Pant, and M. Ram, "7 Security and Privacy challenge in Smart Healthcare and Telemedicine systems," in *Artificial Intelligence, Blockchain and IoT for Smart Healthcare*, River Publishers, 2022, pp. 67–76. Accessed: Jul. 10, 2022. [Online]. Available: https://ieeexplore.ieee.org/document/9782812

[28]    T. H. Szymanski, "Strengthening security and privacy in an ultra-dense green 5G Radio Access Network for the industrial and tactile Internet of Things," in *13th International Wireless Communications and Mobile Computing Conference (IWCMC*, Valencia, 2017, pp. 415–422. doi: 10.1109/IWCMC.2017.7986322.

[29]    K. Demertzis, P. Kikiras, N. Tziritas, S. L. Sanchez, and L. Iliadis, "The Next Generation Cognitive Security Operations Center: Network Flow Forensics Using Cybersecurity Intelligence," *Big Data Cogn. Comput.*, vol. 2, no. 4, Art. no. 4, Dec. 2018, doi: 10.3390/bdcc2040035.

[30]    A. Hassanzadeh, S. Modi, and S. Mulchandani, "Towards effective security control assignment in the Industrial Internet of Things," in *2015 IEEE 2nd World Forum on Internet of Things (WF-IoT*, Milan, 2015, pp. 795–800. doi: 10.1109/WF-IoT.2015.7389155.

[31]    C. Occhiuzzi, S. Amendola, S. Manzari, and G. Marrocco, "Industrial RFID sensing networks for critical infrastructure security," in *46th European Microwave Conference (EuMC*, London, 2016, pp. 1335–1338. doi: 10.1109/EuMC.2016.7824598.

[32]    D. He and S. Zeadally, "An Analysis of RFID Authentication Schemes for Internet of Things in Healthcare Environment Using Elliptic Curve Cryptography," *IEEE Internet Things J.*, vol. 2, no. 1, pp. 72-83, Feb. 2015, doi: 10.1109/JIOT.2014.2360121.

[33] A. Sajid, H. Abbas, and K. Saleem, "Cloud-Assisted IoT-Based SCADA Systems Security: A Review of the State of the Art and Future Challenges," *IEEE Access*, vol. 4, pp. 1375-1384, 2016, doi: 10.1109/ACCESS.2016.2549047.
[34] Z. Drias, A. Serhrouchni, and O. Vogel, "Analysis of cyber security for industrial control systems," in *2015 International Conference on Cyber Security of Smart Cities, Industrial Control System and Communications (SSIC*, Shanghai, 2015, pp. 1–8. doi: 10.1109/SSIC.2015.7245330.
[35] W. Xi and L. Ling, "Research on IoT Privacy Security Risks," in *2016 International Conference on Industrial Informatics - Computing Technology, Intelligent Technology, Industrial Information Integration (ICIICII*, Wuhan, 2016, pp. 259–262. doi: 10.1109/ICIICII.2016.0069.
[36] R. H.Weber, "Internet of things: Privacy issues revisited," *Computer Law & Security Review*, vol. 31, no. ue 5. Elsevier, 2015.
[37] A. R. Sadeghi, C. Wachsmann, and M. Waidner, "Security and privacy challenges in industrial Internet of Things," in *52nd ACM/EDAC/IEEE Design Automation Conference (DAC*, San Francisco, CA, 2015, pp. 1–6. doi: 10.1145/2744769.2747942.
[38] L. D. Xu, W. He, and S. Li, "Internet of Things in Industries: A Survey," *IEEE Trans. Ind. Inform.*, vol. 10, no. 4, pp. 2233-2243, Nov. 2014, doi: 10.1109/TII.2014.2300753.
[39] J. Vávra and M. Hromada, "An evaluation of cyber threats to industrial control systems," in *International Conference on Military Technologies (ICMT) 2015*, Brno, 2015, pp. 1–5. doi: 10.1109/MILTECHS.2015.7153700.
[40] E. C. Strinati *et al.*, "The Hardware Foundation of 6G: The NEW-6G Approach," in *2022 Joint European Conference on Networks and Communications & 6G Summit (EuCNC/6G Summit)*, Jun. 2022, pp. 423–428. doi: 10.1109/EuCNC/6GSummit54941.2022.9815700.
[41] S. Shukla, S. Thakur, and J. G. Breslin, "Secure Communication in Smart Meters using Elliptic Curve Cryptography and Digital Signature Algorithm," in *2021 IEEE International Conference on Cyber Security and Resilience (CSR)*, Jul. 2021, pp. 261–266. doi: 10.1109/CSR51186.2021.9527947.
[42] K. Demertzis, K. Tsiknas, D. Takezis, C. Skianis, and L. Iliadis, "Darknet Traffic Big-Data Analysis and Network Management to Real-Time Automating the Malicious Intent Detection Process by a Weight Agnostic Neural Networks Framework," Feb. 2021, doi: 10.20944/preprints202102.0404.v1.
[43] K. Al Jallad, M. Aljnidi, and M. S. Desouki, "Anomaly detection optimization using big data and deep learning to reduce false-positive," *J. Big Data*, vol. 7, no. 1, p. 68, Aug. 2020, doi: 10.1186/s40537-020-00346-1.
[44] L. Xing, K. Demertzis, and J. Yang, "Identifying data streams anomalies by evolving spiking restricted Boltzmann machines," *Neural Comput. Appl.*, vol. 32, no. 11, pp. 6699–6713, Jun. 2020, doi: 10.1007/s00521-019-04288-5.
[45] S. T. Arzo, R. Bassoli, F. Granelli, and F. H. P. Fitzek, "Multi-Agent Based Autonomic Network Management Architecture," *IEEE Trans. Netw. Serv. Manag.*, vol. 18, no. 3, pp. 3595–3618, Sep. 2021, doi: 10.1109/TNSM.2021.3059752.
[46] A. J. Jara, D. Genoud, and Y. Bocchi, "Big Data for Cyber Physical Systems: An Analysis of Challenges, Solutions and Opportunities," in *Eighth International Conference on Innovative Mobile and Internet Services in Ubiquitous Computing*, Birmingham, 2014, pp. 376–380. doi: 10.1109/IMIS.2014.139.
[47] K. Demertzis and L. Iliadis, "Bio-inspired Hybrid Intelligent Method for Detecting Android Malware," in *Knowledge, Information and Creativity Support Systems*, Cham, 2016, pp. 289–304. doi: 10.1007/978-3-319-27478-2_20.

[48] J. H. Ziegeldorf, O. G. Morchon, and K. Wehrle, "Privacy in the Internet of Things: threats and challenges, Security Comm," *Networks*, 2014, doi: 10.1002/sec.795.

[49] M. Langheinrich, "A survey of RFID privacy approaches," *Pers. Ubiquitous Comput*, vol. 13, no. 6, pp. 413–421, 2009, doi: 10.1007/s00779-008-0213-4.

[50] J. Camenisch *et al.*, "Privacy and identity management for everyone," in *Proceedings of the 2005 workshop on Digital identity management, DIM '05, ACM*, 2005, pp. 20-27,. doi: 10.1145/1102486.1102491.

[51] W. Zhang, C. Wang, and T. Feng, "GP^2S: Generic Privacy-Preservation Solutions for Approximate Aggregation of Sensor Data (concise contribution," *Sixth Annu. IEEE Int. Conf. On*, pp. 179-184, 2008, doi: 10.1109/PERCOM.2008.60.

[52] R. Rios, J. Cuellar, and J. Lopez, "Robust Probabilistic Fake Packet Injection for Receiver-Location Privacy in WSN," in *17th European Symposium on Research in Computer Security (ESORICS 2012), LNCS*, 2012, vol. 7459, pp. 163-180,. doi: 10.1007/978-3-642-33167-1.

[53] A. C. F. Chan and C. Castelluccia, "A security framework for privacy-preserving data aggregation in wireless sensor networks," *ACM Trans. Sens. Netw. TOSN*, vol. 7, no. 4, 2011, doi: 10.1145/1921621.1921623.

[54] J. Camenisch and E. Herreweghen, "Design and implementation of the idemix anonymous credential system," in *Proceedings of the 9th ACM conference on Computer and communications security, CCS '02, ACM*, 2002, pp. 21-30,. doi: 10.1145/586110.586114.

[55] T. Ashur *et al.*, "A Privacy-Preserving Device Tracking System Using a Low-Power Wide-Area Network (LPWAN," in *Cryptology and Network Security, #E International Conference, CANS 2017, Lecture Notes in Computer Science*, 2017.

[56] T. Cnudde and S. Nikova, "Securing the PRESENT Block Cipher Against Combined Side-Channel Analysis and Fault Attacks," *IEEE Trans. Very Large Scale Integr. VLSI Syst.*, vol. 25, no. 10, pp. 1-11, 2017.

[57] A. Hamid, I. B. Baba, and W. Sani, "Proposal for the risk management implementation phase in oil field development project by adding value on the refurbishment of critical equipment," *MATEC Web Conf*, vol. 97, 2017, doi: 10.1051/matecconf/20179701067.

[58] T. Motoyama, "Method and system to erase data after expiration or other condition." Ricoh Company, Ltd, Japan.

[59] M. G., T. C., L. D., and S. R, *Big Data Security Intelligence for Healthcare Industry 4.0*. Cham: Springer, 2017.

[60] J. King, J. Stallings, and M. Riaz, "To log, or not to log: using heuristics to identify mandatory log events – a controlled experiment," *Empir Softw. Eng*, vol. 22, p. 2684, 2017, doi: 10.1007/s10664-016-9449-1.

[61] J. Fitzpatrick, M. Dancho, J. M. Higgins, R. W. Ellis, B. Rub, and CN105593943A, "Data storage system with dynamic erase block grouping mechanism and method of operation thereof US"

[62] Y. Cheng, M. Naslund, G. Selander, and E. Fogelström, "Privacy in machine-to-machine communications A state-of-the-art survey," in *2012 IEEE International Conference on Communication Systems (ICCS*, Singapore, 2012, pp. 75–79. doi: 10.1109/ICCS.2012.6406112.

[63] K. Demertzis and L. Iliadis, "Adaptive Elitist Differential Evolution Extreme Learning Machines on Big Data: Intelligent Recognition of Invasive Species," in *Advances in Big Data*, Cham, 2017, pp. 333–345. doi: 10.1007/978-3-319-47898-2_34.

[64] Esfahani, "A Lightweight Authentication Mechanism for M2M Communications in Industrial IoT Environment," *IEEE Internet Things J.*, no. 99, pp. 1–1, doi: 10.1109/JIOT.2017.2737630.

[65] A. Bartoli, J. Hernandez-Serrano, M. Soriano, M. Dohler, A. Kountouris, and D. Barthel, "Secure Lossless Aggregation for Smart Grid M2M Networks," in *2010 First IEEE International Conference on Smart Grid Communications*, Gaithersburg, MD, 2010, pp. 333–338. doi: 10.1109/SMARTGRID.2010.5622063.
[66] H. Nicholson, "CyberReef Solutions Inc Network-based Machine-To-Machine (M2M) Private Networking System, US 20160337784 A1, Cyberreef Solutions Inc, Louisiana, Assignment Of Assignors Interest"
[67] J. Graham, J. Hieb, and J. Naber, "Improving cybersecurity for Industrial Control Systems," in *2016 IEEE 25th International Symposium on Industrial Electronics (ISIE*, Santa Clara, CA, 2016, pp. 618–623. doi: 10.1109/ISIE.2016.7744960.
[68] E. E. K, J. E, A. L, and M. B, "A Systematic Review of Re-Identification Attacks on Health Data," *PLoS ONE*, vol. 6, no. 12, p. 0028071, 2011, doi: doi:10.1371/.
[69] A. Varghese and A. K. Bose, "Threat modelling of industrial controllers: A firmware security perspective," in *International Conference on Anti-Counterfeiting, Security and Identification (ASID), Macao*, 2014, pp. 1–4. doi: 10.1109/ICASID.2014.7064951.
[70] A. Colella, A. Castiglione, and C. M. Colombini, "Industrial Control System Cyber Threats Indicators in Smart Grid Technology," in *17th International Conference on Network-Based Information Systems*, Salerno, 2014, pp. 374–380. doi: 10.1109/NBiS.2014.129.
[71] A. Razzaq, A. Hur, H. F. Ahmad, and M. Masood, "Cyber security: Threats, reasons, challenges, methodologies and state of the art solutions for industrial applications," in *2013 IEEE Eleventh International Symposium on Autonomous Decentralized Systems (ISADS*, Mexico City, Mexico, 2013, pp. 1–6. doi: 10.1109/ISADS.2013.6513420.
[72] "GSM Association Official Document CLP.13 - IIoT Security Guidelines Endpoint Ecosystem," *Version*, vol. 1, no. 0, Feb. 2016.
[73] A. Hristova, S. Obermeier, and R. Schlegel, "Secure design of engineering software tools in Industrial Automation and Control Systems," in *11th IEEE International Conference on Industrial Informatics (INDIN*, Bochum, 2013, pp. 695–700. doi: 10.1109/INDIN.2013.6622968.
[74] H. Chang, J. Kang, H. Kwon, and C. Lee, "A Research Design on Technology Development for Securing Industrial Assets," in *2nd International Conference on Information Technology Convergence and Services*, Cebu, 2010, pp. 1–4. doi: 10.1109/ITCS.2010.5581273.
[75] L. Edwards, D. McAuley, and L. Diver, "From Privacy Impact Assessment to Social Impact Assessment," in *2016 IEEE Security and Privacy Workshops (SPW*, San Jose, CA, 2016, pp. 53–57. doi: 10.1109/SPW.2016.19.
[76] P. Wang, A. Ali, and W. Kelly, "Data security and threat modeling for smart city infrastructure," in *2015 International Conference on Cyber Security of Smart Cities, Industrial Control System and Communications (SSIC*, Shanghai, 2015, pp. 1–6. doi: 10.1109/SSIC.2015.7245322.
[77] G. Bebrov, R. Dimova, and E. Pencheva, "Quantum approach to the information privacy in Smart Grid," in *International Conference on Optimization of Electrical and Electronic Equipment (OPTIM) & 2017 Intl Aegean Conference on Electrical Machines and Power Electronics (ACEMP*, Brasov, 2017, pp. 971–976. doi: 10.1109/OPTIM.2017.7975096.
[78] M. Rudolph, D. Feth, J. Doerr, and J. Spilker, "Requirements Elicitation and Derivation of Security Policy Templates—An Industrial Case Study," in *2016 IEEE 24th International Requirements Engineering Conference (RE*, Beijing, 2016, pp. 283–292. doi: 10.1109/RE.2016.22.
[79] N. Fabiano, "The Internet of Things ecosystem: The blockchain and privacy issues. The challenge for a global privacy standard," in *2017 International Conference on Internet of

*Things for the Global Community (IoTGC*, Funchal, 2017, pp. 1–7. doi: 10.1109/IoTGC.2017.8008970.
[80]  B. D. Darshini, A. Paventhan, H. Krishna, and N. Pahuja, "Enabling real time requirements in industrial IoT through IETF 6TiSCH," in *2016 International Conference on Internet of Things and Applications (IOTA*, Pune, 2016, pp. 121–124. doi: 10.1109/IOTA.2016.7562707.
[81]  S. M. R. Islam, D. Kwak, M. H. Kabir, M. Hossain, and K. S. Kwak, "The Internet of Things for Health Care: A Comprehensive Survey," *IEEE Access*, vol. 3, pp. 678-708, 2015, doi: 10.1109/ACCESS.2015.2437951.
[82]  A. Borisov, "A Novel Approach for User Authentication to Industrial Components Using QR Codes," in *2015 IEEE 39th Annual Computer Software and Applications Conference*, Taichung, 2015, pp. 61–66. doi: 10.1109/COMPSAC.2015.214.
[83]  T. Song, R. Li, B. Mei, J. Yu, X. Xing, and X. Cheng, "A Privacy Preserving Communication Protocol for IoT Applications in Smart Homes," *IEEE Internet Things J.*, no. 99, pp. 1–1, doi: 10.1109/JIOT.2017.2707489.
[84]  M. Liyanage, M. Ylianttila, and A. Gurtov, "Secure Hierarchical VPLS Architecture for Provider Provisioned Networks," in *IEEE Access*, vol. 3, 2015, pp. 967-984,. doi: 10.1109/ACCESS.2015.2447014.
[85]  A. Arasu, K. Eguro, M. Joglekar, R. Kaushik, D. Kossmann, and R. Ramamurthy, "Transaction processing on confidential data using cipherbase," in *2015 IEEE 31st International Conference on Data Engineering*, Seoul, 2015, pp. 435–446. doi: 10.1109/ICDE.2015.7113304.
[86]  T. H. Chen, W. Shang, A. E. Hassan, M. Nasser, and P. Flora, "Detecting Problems in the Database Access Code of Large Scale Systems - An Industrial Experience Report," in *2016 IEEE/ACM 38th International Conference on Software Engineering Companion (ICSE-C*, Austin, TX, 2016, pp. 71–80.
[87]  O. Cinar, R. H. Guncer, and A. Yazici, "Database Security in Private Database Clouds," in *2016 International Conference on Information Science and Security (ICISS*, Pattaya, 2016, pp. 1–5. doi: 10.1109/ICISSEC.2016.7885847.
[88]  C. Lesjak, H. Bock, D. Hein, and M. Maritsch, "Hardware-secured and transparent multi-stakeholder data exchange for industrial IoT," in *2016 IEEE 14th International Conference on Industrial Informatics (INDIN*, Poitiers, 2016, pp. 706–713. doi: 10.1109/INDIN.2016.7819251.
[89]  S. Katsikeas, "Lightweight & secure industrial IoT communications via the MQ telemetry transport protocol," in *2017 IEEE Symposium on Computers and Communications (ISCC), Heraklion*, Greece, 2017, pp. 1193–1200. doi: 10.1109/ISCC.2017.8024687.
[90]  T. Bartman and K. Carson, "Securing communications for SCADA and critical industrial systems," in *69th Annual Conference for Protective Relay Engineers (CPRE*, College Station, TX, 2016, pp. 1–10. doi: 10.1109/CPRE.2016.7914914.
[91]  M. Ma, D. He, N. Kumar, K. K. R. Choo, and J. Chen, "Certificateless Searchable Public Key Encryption Scheme for Industrial Internet of Things," *IEEE Trans. Ind. Inform.*, no. 99, pp. 1–1, doi: 10.1109/TII.2017.2703922.
[92]  J. Gowthami, N. Shanthi, and N. Krishnamoorthy, "Secure Three-Factor Remote user Authentication for E-Governance of Smart Cities," in *2018 International Conference on Current Trends towards Converging Technologies (ICCTCT)*, Mar. 2018, pp. 1–8. doi: 10.1109/ICCTCT.2018.8551172.
[93]  J. Rrushi and P. A. Nelson, "Big Data Computing for Digital Forensics on Industrial Control Systems," in *2015 IEEE International Conference on Information Reuse and Integration*, San Francisco, CA, 2015, pp. 593–608. doi: 10.1109/IRI.2015.94.

[94]   T. Spyridopoulos, T. Tryfonas, and J. May, "Incident analysis & digital forensics in SCADA and industrial control systems," in *8th IET International System Safety Conference incorporating the Cyber Security Conference 2013*, Cardiff, 2013, pp. 1–6. doi: 10.1049/cp.2013.1720.

[95]   S. Adepu and A. Mathur, "Distributed Attack Detection in a Water Treatment Plant: Method and Case Study," *IEEE Trans. Dependable Secure Comput.*, vol. 18, no. 1, pp. 86–99, Jan. 2021, doi: 10.1109/TDSC.2018.2875008.

[96]   L. Duan, "Automated Policy Combination for Secure Data Sharing in Cross-Organizational Collaborations," in *IEEE Access*, vol. 4, 2016, pp. 3454-3468,. doi: 10.1109/ACCESS.2016.2585185.

[97]   D. Thilakanathan, S. Chen, S. Nepal, and R. Calvo, "SafeProtect: Controlled Data Sharing With User-Defined Policies in Cloud-Based Collaborative Environment," *IEEE Trans. Emerg. Top. Comput.*, vol. 4, no. 2, pp. 301-315, Apr. 2016, doi: 10.1109/TETC.2015.2502429.

[98]   M. Hummer, M. Kunz, M. Netter, L. Fuchs, and G. Pernul, "Advanced Identity and Access Policy Management Using Contextual Data," in *10th International Conference on Availability, Reliability and Security*, Toulouse, 2015, pp. 40–49. doi: 10.1109/ARES.2015.40.

[99]   H. Ryu, D. Kang, and D. Won, "On a Partially Verifiable Multi-party Multi-argument Zero-knowledge Proof," in *2021 15th International Conference on Ubiquitous Information Management and Communication (IMCOM)*, Jan. 2021, pp. 1–8. doi: 10.1109/IMCOM51814.2021.9377407.

[100]   J. Bringer, H. Chabanne, and A. Patey, "Privacy-Preserving Biometric Identification Using Secure Multiparty Computation: An Overview and Recent Trends," *IEEE Signal Process. Mag.*, vol. 30, no. 2, pp. 42–52, Mar. 2013, doi: 10.1109/MSP.2012.2230218.

[101]   B. Jiang, "Two-Party Secure Computation for Any Polynomial Function on Ciphertexts under Different Secret Keys," *Secur. Commun. Netw.*, vol. 2021, p. e6695304, Feb. 2021, doi: 10.1155/2021/6695304.